\documentclass[reprint,showpacs,preprintnumbers,superscriptaddress,nofootinbib,amsmath,amssymb,aps,prl,floatfix]{article}
\pdfoutput=1
\usepackage{jcappub}
\usepackage{graphicx}
\usepackage{dcolumn}
\usepackage{bm}
\usepackage{url}
\usepackage[ddmmyy,24hr]{datetime}
\usepackage{varwidth}
\usepackage{multirow}
\usepackage[center]{subfigure}
\usepackage{xspace}
\usepackage{marginnote}
\usepackage{color}
\newcommand{\ns}{\Delta{N}_{\text{eff}}}
\newcommand{\e}[1]{\times 10^{#1}}
\newcommand{\mpcinv}{\, \text{Mpc}^{-1}}
\newcommand{\eV}{\, \text{eV}}
\newcommand{\cosmomc}{\texttt{CosmoMC}\xspace}
\newcommand{\pchip}{\texttt{PCHIP}\xspace}
\begin{document}


\title{Light Sterile Neutrinos and Inflationary Freedom}

\author[a,b]{S. Gariazzo}

\author[b]{C. Giunti}

\author[c]{M. Laveder}

\affiliation[a]{Department of Physics, University of Torino, Via P. Giuria 1, I--10125 Torino, Italy}
\affiliation[b]{INFN, Sezione di Torino, Via P. Giuria 1, I--10125 Torino, Italy}
\affiliation[c]{Dipartimento di Fisica e Astronomia ``G. Galilei'', Universit\`a di Padova,
and
INFN, Sezione di Padova,
Via F. Marzolo 8, I--35131 Padova, Italy}


\emailAdd{gariazzo@to.infn.it, giunti@to.infn.it, laveder@pd.infn.it}

\abstract{
We perform a cosmological analysis in which we allow the primordial power spectrum of scalar perturbations to assume a shape that is different from the usual power-law
predicted by the simplest models of cosmological inflation.
We parameterize the free primordial power spectrum with a
``piecewise cubic Hermite interpolating polynomial''
(\pchip).
We consider a 3+1 neutrino mixing model with a sterile neutrino
having a mass at the eV scale,
which can explain the anomalies observed in short-baseline neutrino oscillation experiments.
We find that the freedom of the primordial power spectrum
allows to reconcile the cosmological data with a fully thermalized sterile neutrino
in the early Universe.
Moreover, the cosmological analysis gives us some information
on the shape of the primordial power spectrum,
which presents a feature around the wavenumber $k=0.002\mpcinv$.
}


\maketitle

\section{Introduction}

In typical analyses of cosmological data
one of the main assumptions about the early Universe is the form of the primordial power-spectrum (PPS) of scalar fluctuations.
The PPS is usually assumed to be a power-law (PL),
as predicted by the simplest models of inflation
(see Refs.~\cite{Lyth:1998xn,Bassett:2005xm,Baumann:2008bn}).
However,
if inflation is generated by a more complicated mechanism, the PPS can assume a different shape or present various features with respect to the power-law form
(see Refs.~\cite{Martin:2014vha,Kitazawa:2014dya} and references therein).
Since we cannot test directly the physics at the scale of cosmological inflation
in order to check the correctness of the simplest inflationary models,
any cosmological analysis performed assuming a power-law PPS can suffer of a biased constraint.

The cosmological observable that we can access is the late-time power spectrum
of scalar perturbations,
which is a convolution of the PPS and the transfer function,
that can be calculated numerically as a function of a small number of cosmological parameters.
Several experiments are designed to measure the late-time power spectrum at different redshifts
(see Refs.~\cite{Tegmark:2003ud,Parkinson:2012vd,Ahn:2012fh}).

The physics of the transfer function is well understood and the experiments that measure the Cosmic Microwave Background (CMB) radiation give very strong constraints on the values of the cosmological parameters which determine the transfer function.
The current most precise measurements of the angular power spectrum of the CMB are those of the Planck experiment \cite{Ade:2013ktc} for the unpolarized data and those of
the WMAP experiment \cite{Bennett:2012zja} for the polarization spectra.
However,
the next Planck data release is expected to improve the current sensitivity on the unpolarized spectra and to include the new polarized spectra obtained by Planck.

On the other hand,
since the inflationary scale cannot be directly tested,
we can only try to reconstruct indirectly the PPS.
In the literature there are several approaches for reconstructing a
completely unknown PPS given the available experimental data.
Among them we can list
the ``cosmic inversion'' methods \cite{Matsumiya:2001xj,Matsumiya:2002tx,Kogo:2003yb,Kogo:2005qi,Nagata:2008tk},
maximum entropy deconvolution \cite{Goswami:2013uja}
and regularization methods like
Richardson-Lucy iteration \cite{Shafieloo:2003gf,Nicholson:2009pi,Hazra:2013eva,Hazra:2014jwa},
truncated singular value decomposition \cite{Nicholson:2009zj}
and Tikhonov regularization \cite{Hunt:2013bha}.

The effects on cosmological parameter estimation of considering a PPS which is different from a power-law
has been studied in several works: for example, the power-law PPS has been simply modified with the introduction of
a running in the tilt of the power-law \cite{Archidiacono:2012gv,Zhao:2012xw,Vazquez:2013dva,Abazajian:2014tqa},
a running of the running \cite{Cheng:2014pna},
or
a sharp cut-off in the power-law \cite{Abazajian:2014tqa}.
Our main goal is to study how the freedom of the form of the PPS can affect the existing bounds on the presence in the early Universe of additional sterile neutrinos.
In particular,
we want to explore the impact of a light sterile neutrino with mass $m_s\sim1\eV$
which has been thermalized by neutrino oscillations before
neutrino decoupling at a temperature of the order of 1 MeV
\cite{Dolgov:2003sg,Cirelli:2004cz}.
Previous analyses of the cosmological data
with a standard power-law PPS have found that
the case of a fully thermalized sterile neutrino is quite disfavored
\cite{Valentino:2013wha,Archidiacono:2013xxa,Mirizzi:2013kva,Gariazzo:2013gua,Archidiacono:2014apa,Bergstrom:2014fqa}.
This result motivated the study of mechanisms which can suppress the
thermalization of sterile neutrinos in the early Universe
due to active-sterile oscillations before neutrino decoupling
\cite{Dolgov:2003sg,Cirelli:2004cz}.
Examples are a large lepton asymmetry
\cite{Chu:2006ua,Hannestad:2012ky,Mirizzi:2012we,Saviano:2013ktj,Hannestad:2013pha},
an enhanced background potential due to new interactions in the sterile sector
\cite{Hannestad:2013ana,Dasgupta:2013zpn,Bringmann:2013vra,Ko:2014bka,Archidiacono:2014nda,Saviano:2014esa,Mirizzi:2014ama},
a larger cosmic expansion rate at the time of sterile neutrino production
\cite{Rehagen:2014vna},
and
MeV dark matter annihilation
\cite{Ho:2012br}.

Besides our main objective,
which is to find out how the constraints on the
presence in the early Universe of additional sterile neutrinos
change if the PPS is not forced to be a power-law,
we are also interested in obtaining information on the form of the PPS.
With these aims,
we considered a general form of the PPS that allows the presence of features without forcing a particular shape.
In the literature several model-independent parameterizations have been used:
for example,
a free PPS can be described with
wavelets \cite{Mukherjee:2003cz,Mukherjee:2003yx,Mukherjee:2003ag,Mukherjee:2005dc},
principal components \cite{Leach:2005av},
top-hat bins without interpolation \cite{Wang:1998gb},
power-law bins \cite{Hannestad:2000pm,Hazra:2013nca},
linear interpolation
\cite{Wang:2000js,Bridle:2003sa,Hannestad:2003zs,Bridges:2005br,Spergel:2006hy,Bridges:2006zm,Bridges:2008ta,Vazquez:2013dva},
broken power-law \cite{Hazra:2013nca,Hazra:2014aea},
and
interpolating spline functions
\cite{Sealfon:2005em,Verde:2008zza,Peiris:2009wp,Hlozek:2011pc,Gauthier:2012aq,dePutter:2014hza,Abazajian:2014tqa,Hu:2014aua}.
We decided to follow part of the prescriptions of the interpolating spline form presented in Refs.~\cite{Hlozek:2011pc,dePutter:2014hza},
improving the parametrization by using a
``piecewise cubic Hermite interpolating polynomial''
(\pchip),
which is described in Appendix~\ref{sec:ppsparametrization}.
This method
allows us to avoid the spurious oscillating behavior that can appear
between the nodes of interpolating splines.

This article is structured as follows:
in Sec.~\ref{sec:cosmoanalysis} we introduce the neutrino mixing scheme, the cosmological model and the cosmological data used in the paper,
in Sec.~\ref{sec:results_cosmo} and
in Sec.~\ref{sec:results_PPS} we discuss the results obtained from the analysis for the cosmological parameters and for the PPS respectively,
and
in Sec.~\ref{sec:disc} we present our conclusions.

\section{Neutrino mixing scheme, cosmological model and data}
\label{sec:cosmoanalysis}

In this Section we introduce the models and the datasets used in this paper.
In Subsection~\ref{sub:numix} we present the neutrino mixing scheme,
in Subsection~\ref{sub:cosmomodel}
we introduce the cosmological model, and
in Subsection~\ref{sub:cosmodata}
we present the cosmological data used in our analysis.

\subsection{Neutrino mixing scheme}
\label{sub:numix}

In this paper we consider the 3+1 neutrino mixing scheme,
which is motivated by indications in favor of short-baseline
neutrino oscillations
found in the LSND experiment
\cite{Aguilar:2001ty},
in Gallium experiments
\cite{Abdurashitov:2005tb,Laveder:2007zz,Giunti:2006bj,Acero:2007su,Giunti:2010zu}
and in reactor experiments
\cite{Mueller:2011nm,Mention:2011rk,Huber:2011wv}.
In this scheme,
besides the three standard active flavor neutrinos
$\nu_{e}$,
$\nu_{\mu}$,
$\nu_{\tau}$,
there is a sterile neutrino
which does not interact through standard weak interactions.
This sterile neutrino is a new particle beyond the Standard Model which cannot be detected directly in current experiments
(see \cite{Bilenky:1998dt,GonzalezGarcia:2007ib,Abazajian:2012ys}).

The four flavor neutrinos
$\nu_{e}$,
$\nu_{\mu}$,
$\nu_{\tau}$,
$\nu_{s}$
are superpositions of four massive neutrinos,
$\nu_{1}$,
$\nu_{2}$,
$\nu_{3}$,
$\nu_{4}$
with respective masses
$m_{1}$,
$m_{2}$,
$m_{3}$,
$m_{4}$.
The squared mass differences
$
\Delta{m}^{2}_{21}
\simeq
8 \times 10^{-5} \eV^{2}
$
and
$
\Delta{m}^{2}_{31}
\simeq
2 \times 10^{-3} \eV^{2}
$
(where
$\Delta{m}^{2}_{kj} = m_{k}^2 - m_{j}^2$)
generate the neutrino oscillations which have been observed
in many solar, atmospheric and long-baseline experiments
(see \cite{GonzalezGarcia:2012sz,Bellini:2013wra,Capozzi:2013csa,NuFIT}).
An additional much larger squared mass difference
$
\Delta{m}^{2}_{41}
\sim
1 \, \eV^{2}
$
is required
in order to explain the indications in favor of short-baseline oscillations
\cite{Kopp:2011qd,Giunti:2011gz,Giunti:2011hn,Giunti:2011cp,Conrad:2012qt,Kopp:2013vaa,Giunti:2013aea}.
In the 3+1 scheme
the three standard active flavor neutrinos
$\nu_{e}$,
$\nu_{\mu}$,
$\nu_{\tau}$
are mainly composed of the three massive neutrinos
$\nu_{1}$,
$\nu_{2}$,
$\nu_{3}$,
but they have a small component of $\nu_{4}$ in order to generate the observed short-baseline oscillations
through $\Delta{m}^{2}_{41}$.
On the other hand,
the sterile neutrino $\nu_{s}$ is mainly composed of
the massive neutrino $\nu_4$
and
in the following we use the common notation $m_{s}=m_{4}$.

Since the case of a very light
$\nu_{4}$
and almost degenerate
$\nu_{1}$,
$\nu_{2}$,
$\nu_{3}$
at the eV scale is strongly disfavored by cosmological data
(see Ref.~\cite{Ade:2013zuv})
we consider the case of
$m_{s}=m_{4}\sim1\,\text{eV}$
and much lighter
$\nu_{1}$,
$\nu_{2}$,
$\nu_{3}$.

The combined analysis of cosmological data and
short-baseline neutrino oscillation data is performed
by using the posterior distribution of
$m_s = m_4 \simeq \sqrt{\Delta{m}^2_{41}}$
obtained from the analysis of SBL data
\cite{Giunti:2013aea}
as a prior in the \cosmomc analysis of cosmological data
\cite{Archidiacono:2012ri,Archidiacono:2013xxa,Gariazzo:2013gua,Archidiacono:2014apa,Gariazzo:2014pja}.
As shown in Tab.~3 of Ref.~\cite{Gariazzo:2013gua},
the best-fit value of $m_s$ obtained from short-baseline neutrino oscillation data
is 1.27 eV
and its 95.45\% probability range ($2\sigma$)
is between
0.97 and 1.42 eV.

\subsection{Cosmological model}
\label{sub:cosmomodel}

We used an extended flat $\Lambda$CDM model to accommodate the presence of a sterile neutrino and inflationary freedom in the production of the primordial power spectra.

In the analysis with a power-law PPS we consider a flat
$\Lambda$CDM+$\nu_s$ cosmological model with a total of
eight parameters:
\begin{equation}\label{eq:modelbase}
{\bm \theta}
=
\{\omega_{\rm cdm},\omega_{\rm b},\theta_{\rm s},\tau,
\ln(10^{10}A_{s}),n_{s},m_s,\ns
\}
,
\end{equation}
where
$\omega_{\rm cdm} \equiv \Omega_{\rm cdm} h^2$ and $\omega_{\rm b} \equiv \Omega_{\rm b} h^2$ are the present-day physical CDM and baryon densities,
$\theta_{\rm s}$ the angular the sound horizon,
$\tau$  the optical depth to reionisation, and
$\ln(10^{10}A_{s})$ and $n_s$ denote respectively the amplitude and spectral index of the initial scalar fluctuations 
at the pivot scale of 0.002 Mpc$^{-1}$.
$\ns = N_{\text{eff}} - N_{\text{eff}}^{\text{SM}}$,
where
$N_{\text{eff}}^{\text{SM}} = 3.046$ \cite{Mangano:2005cc}
is the effective number of relativistic degrees of freedom
before photon decoupling in the Standard Model with three massless neutrinos
(see \cite{Archidiacono:2013fha,Lesgourgues:2014zoa}).

In contrast with previous analyses
\cite{Gariazzo:2013gua,Archidiacono:2014apa,Gariazzo:2014pja},
we limit the allowed range of
$\ns$ in the interval $0\leq\ns\leq1$,
assuming that the additional sterile neutrino cannot contribute to the relativistic energy density more than a standard active neutrino.
This happens if sterile neutrinos are produced in the early Universe by
neutrino oscillations before neutrino decoupling
\cite{Dolgov:2003sg,Cirelli:2004cz}.

We assume a flat prior for all the parameters in Eq.~(\ref{eq:modelbase}),
except $m_s$,
for which we use a flat prior
for $0\leq m_s/\eV\leq 3$ only in the analyses which do not take into account the constraints
from short-baseline neutrino oscillation data.
In the analyses which take into account these constraints
we use as prior for $m_s$ the posterior obtained from the analysis of SBL data presented in Ref.~\cite{Giunti:2013aea},
as explained at the end of Subsection~\ref{sub:numix}.
We neglect the masses of the three light neutrinos
$\nu_{1}$,
$\nu_{2}$,
$\nu_{3}$,
which are assumed to be much smaller than 1 eV.

In order to parameterize a free PPS we follow partially the prescriptions described in \cite{Hlozek:2011pc,dePutter:2014hza},
but instead of the cubic spline function we interpolate with a
``piecewise cubic Hermite interpolating polynomial'' (\pchip)
\cite{Fritsch:1980,Fritsch:1984},
that is described only by the values of the PPS in a discrete number of nodes,
as discussed in Appendix~\ref{sec:ppsparametrization}.
We used 12 nodes
which span a wide range of $k$ values:
\begin{align}
k_1     &= 5\e{-6} \mpcinv , \nonumber\\
k_2     &= 10^{-3} \mpcinv , \nonumber\\
k_j     &= k_2 (k_{11}/k_2)^{(j-2)/9} \quad \text{for} \quad j\in[3,10] , \nonumber\\
k_{11}  &= 0.35 \mpcinv , \nonumber\\
k_{12}  &= 10\mpcinv .
\label{eq:nodesspacing}
\end{align}
In the range $(k_2, k_{11})$, that is well constrained from the data \cite{dePutter:2014hza},
we choose equally spaced nodes in the logarithmic scale.
The nodes $k_1$ and $k_{12}$ are used to parameterize a non-constant behaviour of the PPS outside this range
and their position is chosen in order to have all the
\cosmomc PPS evaluations inside the interval covered by our parametrization.
The \pchip PPS is described by
\begin{equation}
P_{s}(k)=P_0 \times \pchip(k; P_{s,1}, \ldots, P_{s,12})
,
\label{eq:pchip}
\end{equation}
where $P_0=2.36\e{-9}$ \cite{Larson:2010gs}
and $P_{s,j}$ is the value of the PPS at the node $k_j$ divided by $P_0$.
The function $\pchip(k; P_{s,1}, \ldots, P_{s,12})$
is described in Appendix~\ref{sec:ppsparametrization}.

In the \pchip PPS analysis we consider a flat
$\Lambda$CDM+$\nu_s$ cosmological model with a total of
18 parameters:
\begin{equation}\label{eq:modelmodified}
{\bm \theta} = \{\omega_{\rm cdm},\omega_{\rm b}, \theta_{\rm s}, \tau, m_s, \ns,
P_{s,1}, \ldots, P_{s,12}
\}
,
\end{equation}
where $\omega_{\rm cdm},\omega_{\rm b},\theta_{\rm s},\tau,m_s,\ns $ are the same as
those in the set (\ref{eq:modelbase}).
We assume a flat prior on $P_{s,j}$ in the range $0.01\leq P_{s,j}\leq 10$.

The Bayesian analysis is performed through an appropriately modified version of the Monte Carlo Markov Chain (MCMC) package
\cosmomc \cite{Lewis:2002ah},
using the Boltzmann equations solver \texttt{CAMB} \cite{Lewis:1999bs} (Code for Anisotropies in the Microwave Background)
for the calculation of the observables.

\subsection{Cosmological data sets}
\label{sub:cosmodata}

In this paper we use the same dataset as in Refs.~\cite{Archidiacono:2014apa,Gariazzo:2014pja},
apart from the controversial BICEP2 data
on the B-mode polarization of the CMB
\cite{Ade:2014xna}
that we neglect:

\begin{itemize}

\item {\bf CMB} (Cosmic Microwave Radiation):
the temperature fluctuations power spectra provided by the Planck satellite
\cite{Ade:2013kta}
up to $\ell=2479$,
by Atacama Cosmology Telescope (ACT) \cite{Dunkley:2013vu}
and South Pole Telescope (SPT) \cite{Story:2012wx}
whose likelihoods cover the high multipole range,
$500<\ell<3500$ and $650<\ell<3000$, respectively.
Concerning polarization we include the data of the Wilkinson Microwave Anisotropy Probe (WMAP) nine year data release \cite{Bennett:2012zja}.

\item {\bf LSS} (Large Scale Structure):
the matter power spectrum at four different redshifts $z=0.22$, $z=0.41$, $z=0.60$ and $z=0.78$ from the WiggleZ Dark Energy Survey \cite{Parkinson:2012vd}.

\item {$\mathbf{H_0}$}:
the Hubble parameter as obtained with the Hubble Space Telescope (HST) \cite{Riess:2011yx}, which acts as a prior on the derived cosmological parameter $H_0 =73.8\pm2.4 \, \text{km} \, \text{s}^{-1} \, \text{Mpc}^{-1}$.

\item {\bf PSZ}:
The Planck Sunayev Zel'Dovich catalogue \cite{Ade:2013lmv}
contains 189 galaxy clusters identified through the Sunayev Zel'Dovich effect.
The number counts allows to compute the cluster mass function, which is related to a combination of
$\Omega_m$ and $\sigma_8$: $\sigma_8(\Omega_m/0.27)^{0.3}=0.782\pm0.010$. This result contributes as an additional $\chi^2$ in our analysis.

\item {\bf CFHTLenS}:
the 2D cosmic shear correlation function as determined by the Canada-France Hawaii Telescope Lensing Survey (CFHTLenS)
\cite{Kilbinger:2012qz,Heymans:2013fya} through the measurements of redshifts and shapes of 4.2 million galaxies spanning the range $0.2<z<1.3$.
The weak gravitational lensing signal extracted from these measurements constrains a combination of the total matter density and the standard deviation of the amplitude of the matter density fluctuations on a sphere of radius $8h^{-1}{\rm Mpc}$: $\sigma_8(\Omega_m/0.27)^{0.46}=0.774\pm0.040$.
This result is incorporated in our analysis following the same prescription used for PSZ.

\end{itemize}

In the following we denote the analyses of all these cosmological data alone as ``COSMO''
and those which include also the short-baseline neutrino oscillation prior as
``COSMO+SBL''.

\begin{table}
\begin{center}
\renewcommand{\arraystretch}{1.4}
\begin{tabular}{|l|c|c|}
\hline
Parameters
&
COSMO
&
COSMO+SBL
\\

\hline

$100\,\Omega_{\rm b} h^2$ 	
			& $2.263^{+0.026}_{-0.027}\,^{+0.052}_{-0.053}\,^{+0.078}_{-0.080}$        & $2.251^{+0.023}_{-0.025}\,^{+0.049}_{-0.045}\,^{+0.075}_{-0.067}$        \\
                             
$\Omega_{\rm cdm} h^2$       
			& $0.120^{+0.004}_{-0.005}\,^{+0.008}_{-0.008}\,^{+0.011}_{-0.009}$        & $0.117^{+0.002}_{-0.003}\,^{+0.006}_{-0.005}\,^{+0.010}_{-0.006}$        \\
                             
$\theta_{\rm s}$             
			& $1.0412^{+0.0007}_{-0.0007}\,^{+0.0014}_{-0.0014}\,^{+0.0020}_{-0.0021}$ & $1.0416^{+0.0006}_{-0.0006}\,^{+0.0012}_{-0.0012}\,^{+0.0018}_{-0.0019}$ \\
                             
$\tau$                       
			& $0.087^{+0.013}_{-0.014}\,^{+0.028}_{-0.026}\,^{+0.045}_{-0.037}$        & $0.087^{+0.013}_{-0.013}\,^{+0.026}_{-0.025}\,^{+0.040}_{-0.035}$        \\
\hline                       
$\ns$                        
			& $0.38^{+0.18}_{-0.33}$; No limit; No limit                               & $0.19^{+0.09}_{-0.12}$; $<0.41$; $<0.60$                               \\
                                                                                    
$m_s [\rm{eV}]$                                                                     
			& $0.61^{+0.31}_{-0.42}$; $<2.03$; No limit                                & $1.25^{+0.11}_{-0.16}\,^{+0.17}_{-0.29}\,^{+0.22}_{-0.35}$               \\
\hline                             
$n_{\rm s}$                  
			& $0.979^{+0.011}_{-0.010}\,^{+0.020}_{-0.020}\,^{+0.030}_{-0.025}$        & $0.969^{+0.005}_{-0.005}\,^{+0.011}_{-0.011}\,^{+0.017}_{-0.016}$        \\
                             
$\log(10^{10} A_s)$          
			& $3.152^{+0.031}_{-0.032}\,^{+0.064}_{-0.058}\,^{+0.094}_{-0.087}$        & $3.178^{+0.024}_{-0.025}\,^{+0.048}_{-0.051}\,^{+0.072}_{-0.075}$        \\
\hline
\end{tabular}
\end{center}
\caption{\label{tab:plPPS} 
Marginalized $1\sigma$, $2\sigma$ and $3\sigma$ confidence level limits for the cosmological parameters
obtained with the power-law parametrization for the PPS.
}
\end{table}

\begin{table}
\begin{center}
\renewcommand{\arraystretch}{1.4}
\begin{tabular}{|l|c|c|}
\hline
Parameters
&
COSMO
&
COSMO+SBL
\\

\hline

$100\,\Omega_{\rm b} h^2$ 	
			& $2.251^{+0.036}_{-0.036}\,^{+0.073}_{-0.073}\,^{+0.111}_{-0.110}$        & $2.247^{+0.036}_{-0.038}\,^{+0.072}_{-0.078}\,^{+0.111}_{-0.117}$        \\
                        
$\Omega_{\rm cdm} h^2$  
			& $0.125^{+0.005}_{-0.004}\,^{+0.007}_{-0.011}\,^{+0.009}_{-0.014}$        & $0.118^{+0.004}_{-0.005}\,^{+0.011}_{-0.007}\,^{+0.016}_{-0.008}$        \\
                        
$\theta_{\rm s}$        
			& $1.0407^{+0.0007}_{-0.0008}\,^{+0.0016}_{-0.0014}\,^{+0.0024}_{-0.0020}$ & $1.0413^{+0.0008}_{-0.0007}\,^{+0.0014}_{-0.0016}\,^{+0.0020}_{-0.0024}$ \\
                        
$\tau$                  
			& $0.086^{+0.014}_{-0.016}\,^{+0.033}_{-0.028}\,^{+0.053}_{-0.038}$        & $0.090^{+0.014}_{-0.016}\,^{+0.033}_{-0.029}\,^{+0.051}_{-0.039}$        \\
\hline                                                                                                         
$\ns$                                                                                                                              
			& $>0.54$; No limit; No limit                                              & $0.25^{+0.13}_{-0.22}$; $<0.75$; No limit                                \\
                                                                                                                                             
$m_s [\rm{eV}]$                                                                                                                              
			& $0.62^{+0.21}_{-0.26}\,^{+0.87}_{-0.54}$; No limit                       & $1.22^{+0.13}_{-0.15}\,^{+0.17}_{-0.28}\,^{+0.24}_{-0.33}$               \\
\hline
$P_{s,1}$
			& $<2.51$; $<8.13$; No limit                                               & $<2.75$; $<8.30$; No limit                                               \\

$P_{s,2}$
			& $1.06^{+0.19}_{-0.22}\,^{+0.43}_{-0.35}\,^{+0.71}_{-0.43}$               & $1.05^{+0.18}_{-0.22}\,^{+0.44}_{-0.35}\,^{+0.75}_{-0.44}$               \\

$P_{s,3}$
			& $0.65^{+0.19}_{-0.19}\,^{+0.38}_{-0.37}\,^{+0.57}_{-0.54}$               & $0.67^{+0.20}_{-0.19}\,^{+0.39}_{-0.36}\,^{+0.61}_{-0.52}$               \\

$P_{s,4}$
			& $1.14^{+0.11}_{-0.11}\,^{+0.23}_{-0.22}\,^{+0.36}_{-0.31}$               & $1.13^{+0.11}_{-0.11}\,^{+0.23}_{-0.21}\,^{+0.34}_{-0.31}$               \\

$P_{s,5}$
			& $0.97^{+0.06}_{-0.05}\,^{+0.11}_{-0.10}\,^{+0.17}_{-0.16}$               & $0.98^{+0.05}_{-0.06}\,^{+0.11}_{-0.10}\,^{+0.17}_{-0.15}$               \\

$P_{s,6}$
			& $0.96^{+0.03}_{-0.03}\,^{+0.07}_{-0.06}\,^{+0.10}_{-0.08}$               & $0.98^{+0.03}_{-0.03}\,^{+0.07}_{-0.06}\,^{+0.11}_{-0.08}$               \\

$P_{s,7}$
			& $0.94^{+0.03}_{-0.03}\,^{+0.06}_{-0.05}\,^{+0.10}_{-0.07}$               & $0.94^{+0.03}_{-0.03}\,^{+0.06}_{-0.06}\,^{+0.10}_{-0.07}$               \\

$P_{s,8}$
			& $0.93^{+0.03}_{-0.03}\,^{+0.06}_{-0.05}\,^{+0.10}_{-0.07}$               & $0.93^{+0.03}_{-0.03}\,^{+0.06}_{-0.06}\,^{+0.10}_{-0.07}$               \\

$P_{s,9}$
			& $0.93^{+0.03}_{-0.03}\,^{+0.07}_{-0.06}\,^{+0.11}_{-0.08}$               & $0.91^{+0.03}_{-0.03}\,^{+0.07}_{-0.06}\,^{+0.10}_{-0.07}$               \\

$P_{s,10}$
			& $0.91^{+0.04}_{-0.04}\,^{+0.08}_{-0.08}\,^{+0.12}_{-0.11}$               & $0.88^{+0.03}_{-0.04}\,^{+0.08}_{-0.07}\,^{+0.14}_{-0.08}$               \\

$P_{s,11}$
			& $1.14^{+0.17}_{-0.16}\,^{+0.28}_{-0.30}\,^{+0.42}_{-0.39}$               & $1.00^{+0.13}_{-0.17}\,^{+0.35}_{-0.24}\,^{+0.52}_{-0.28}$               \\

$P_{s,12}$
			& $<0.70$; $<1.19$; $<1.54$                                                & $<0.49$; $<1.01$; $<1.33$                                                \\
\hline
\end{tabular}
\end{center}
\caption{\label{tab:freePPS}
Marginalized $1\sigma$, $2\sigma$ and $3\sigma$ confidence level limits for the cosmological parameters
obtained with the \pchip{} parametrization for the PPS.
}
\end{table}

\section{Cosmological Parameters and Sterile Neutrinos}
\label{sec:results_cosmo}

The results of our COSMO and COSMO+SBL analyses are presented in
Tab.~\ref{tab:plPPS} for the standard case of a power-law PPS
and
in Tab.~\ref{tab:freePPS} for the free PPS with the \pchip parameterization.
In the upper part of the tables we listed the common parameters of the $\Lambda$CDM model,
in the central part we listed the neutrino parameters $\ns$ and $m_s$,
while the lower part concerns the parameters used to parameterize the PPS:
$n_{\rm s}$ and $\log(10^{10} A_s)$ for the power-law PPS
and $P_{s,j}$ for the \pchip PPS.
The constraints on the PPS parameters are discussed in the next section.
In this section we discuss first the results relative to the parameters in the upper part of the tables, $100\,\Omega_{\rm b} h^2$, $\Omega_{\rm cdm} h^2$, $\theta_{\rm s}$ and $\tau$,
and then the results relative to the parameters in the central part of the tables,
$\ns$ and $m_s$.

The bounds on the parameters of the $\Lambda$CDM model change slightly when more freedom is admitted for the PPS.
Comparing Tabs.~\ref{tab:plPPS} and \ref{tab:freePPS},
one can see that
the limits
on the parameters of the $\Lambda$CDM model
are slightly weakened in the \pchip PPS case and for some parameters there is also a small shift in the marginalized best-fit value.
In all the cases in which this happens, the marginalized best-fit values move inside the $1\sigma$ uncertainties.
The freedom of the form of the PPS affects the COSMO results
more than the COSMO+SBL results:
in the former case the $\Omega_{\rm cdm} h^2$ and $\theta_{\rm s}$ best values change by about $1\sigma$, while a smaller shift is obtained for $100\,\Omega_{\rm b} h^2$.
On the other hand, in the COSMO+SBL analysis all the shifts are much smaller than the $1\sigma$ uncertainties.

\begin{figure*}
\centering
\includegraphics[width=0.7\textwidth]{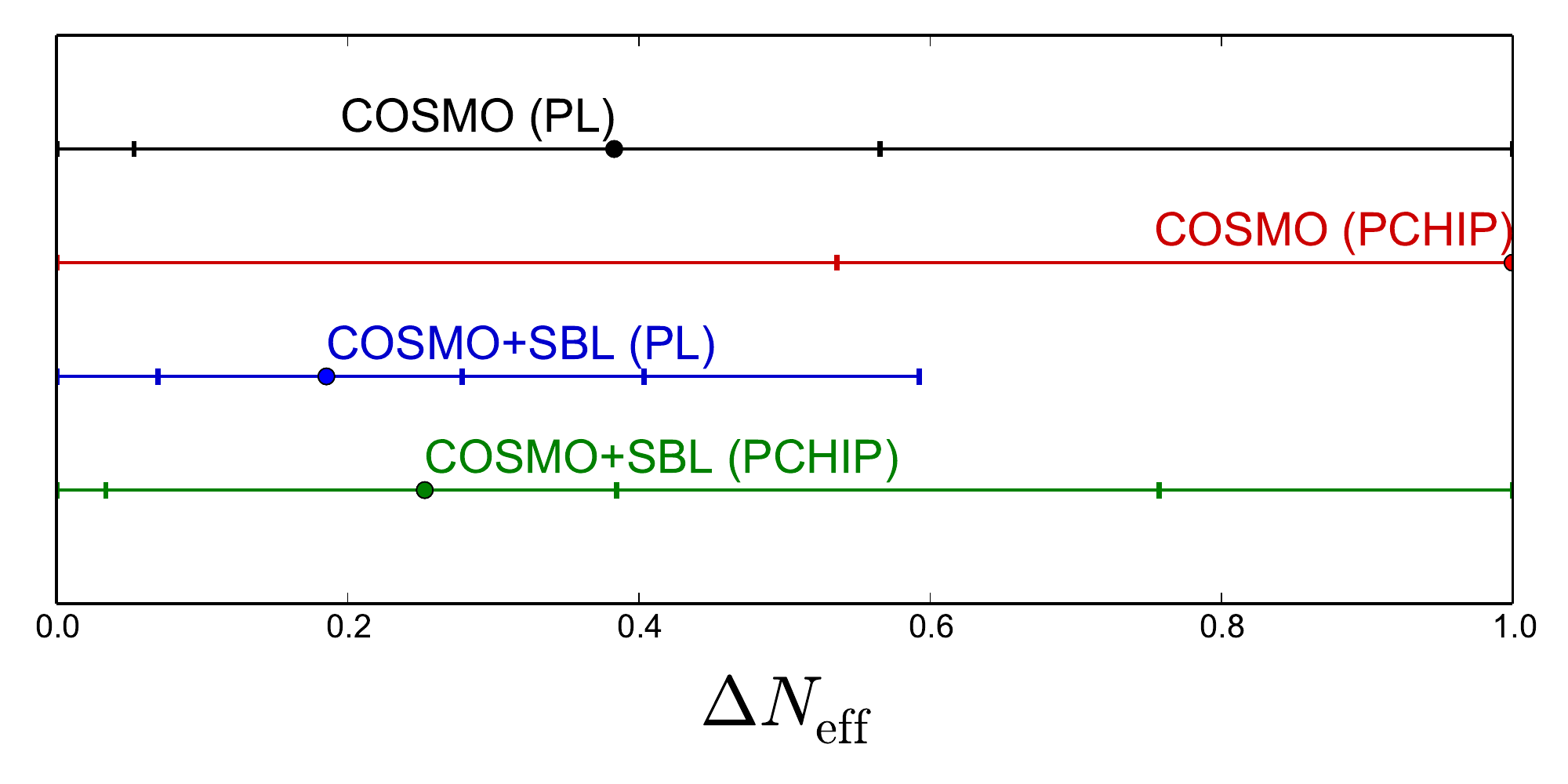}
\includegraphics[width=0.7\textwidth]{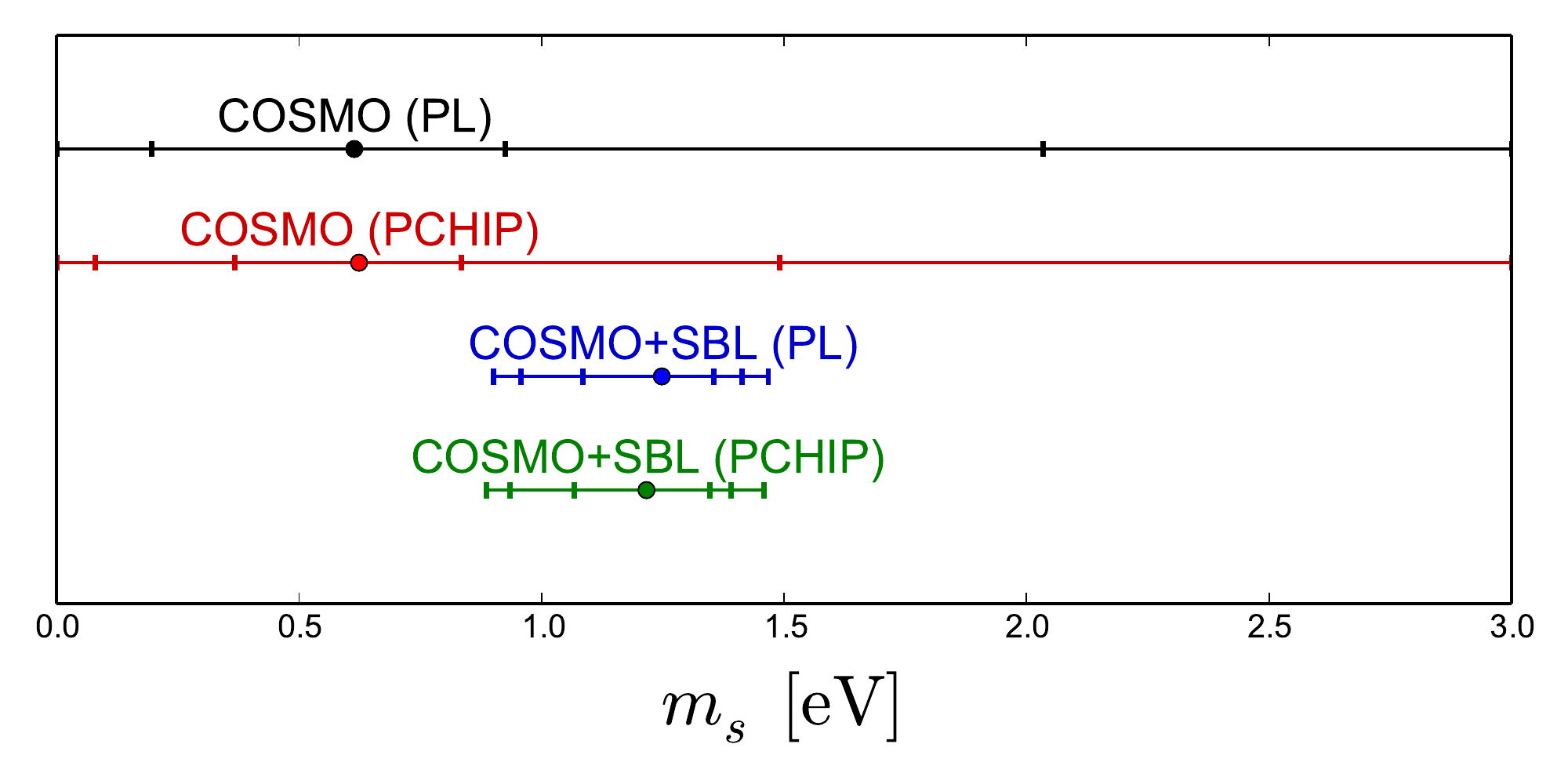}
\caption{\label{fig:errorbars}
$1\sigma$, $2\sigma$ and $3\sigma$ marginalized intervals for $\ns$ and $m_s$
obtained in the different analyses discussed in the text
(considering $0\leq\ns\leq1$ and $0\leq m_s/\eV\leq 3$).
}
\end{figure*}

\begin{figure*}
\centering
\includegraphics[width=0.49\textwidth]{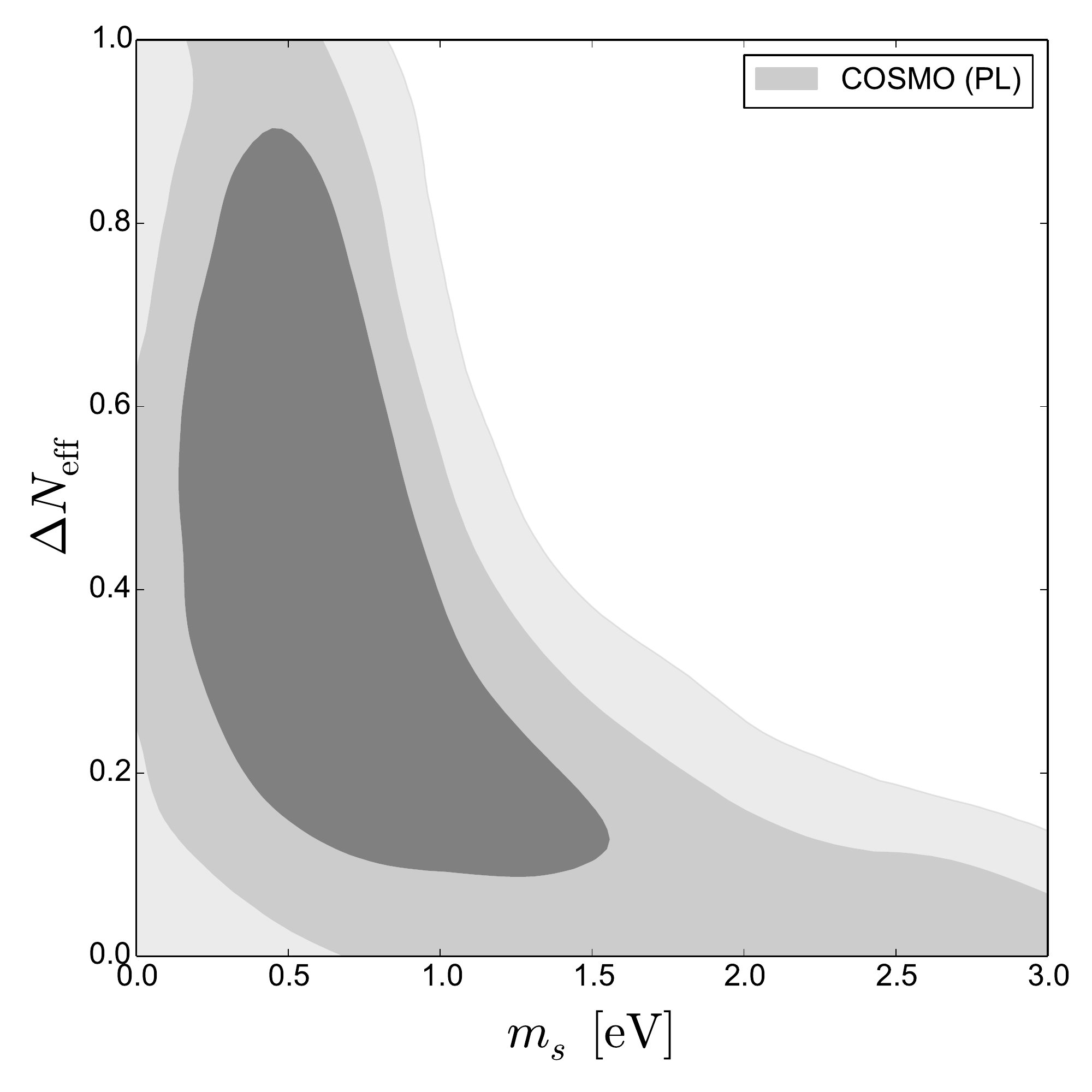}
\includegraphics[width=0.49\textwidth]{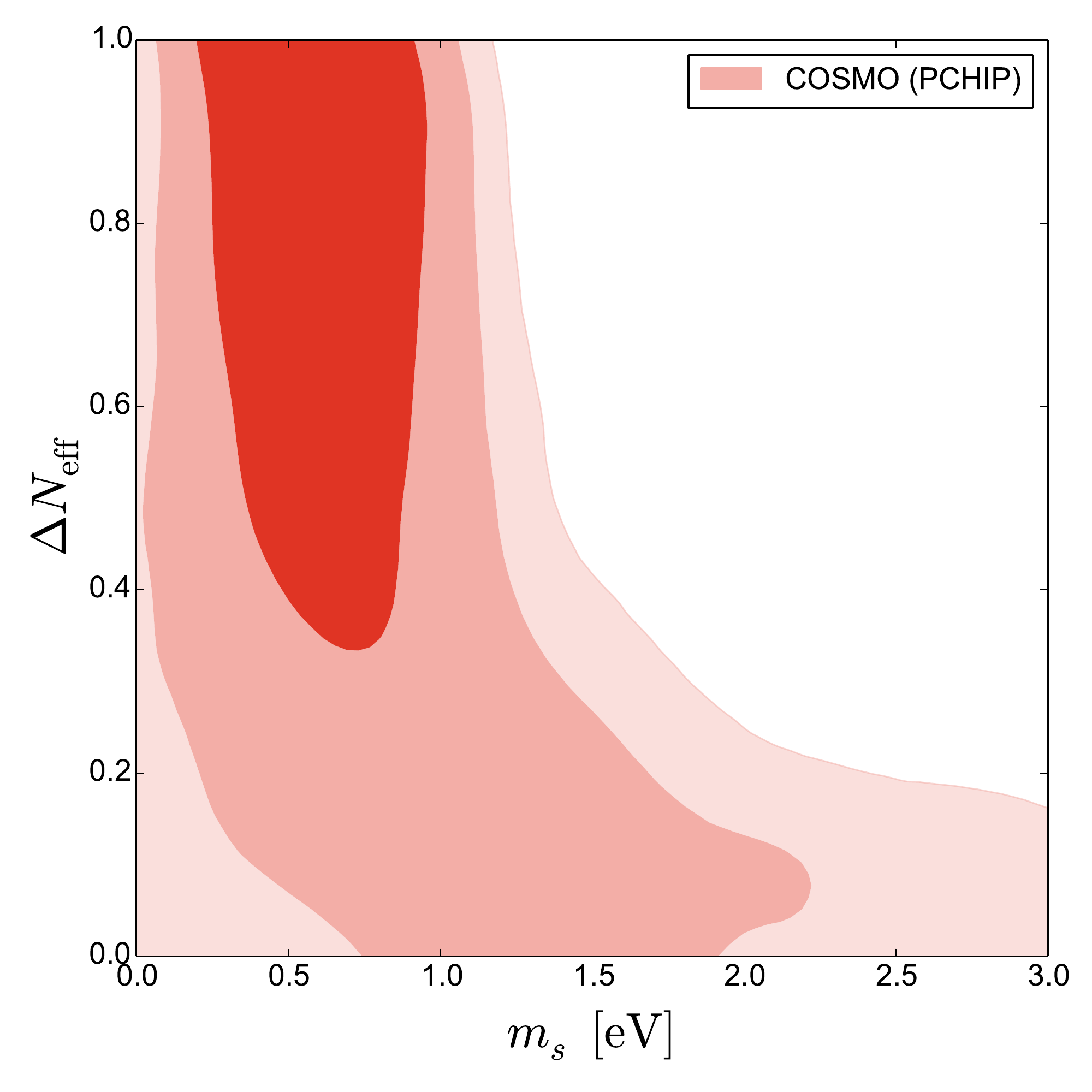}
\caption{\label{fig:posterior2DnoSBL}
$1\sigma$, $2\sigma$ and $3\sigma$ marginalized contours in the $m_s-\ns$ plane
in the fits without the SBL prior.
The left and right panels correspond, respectively,
to the standard power-law PPS and the \pchip PPS analyses.
}
\vspace{0.5cm}
\includegraphics[width=0.49\textwidth]{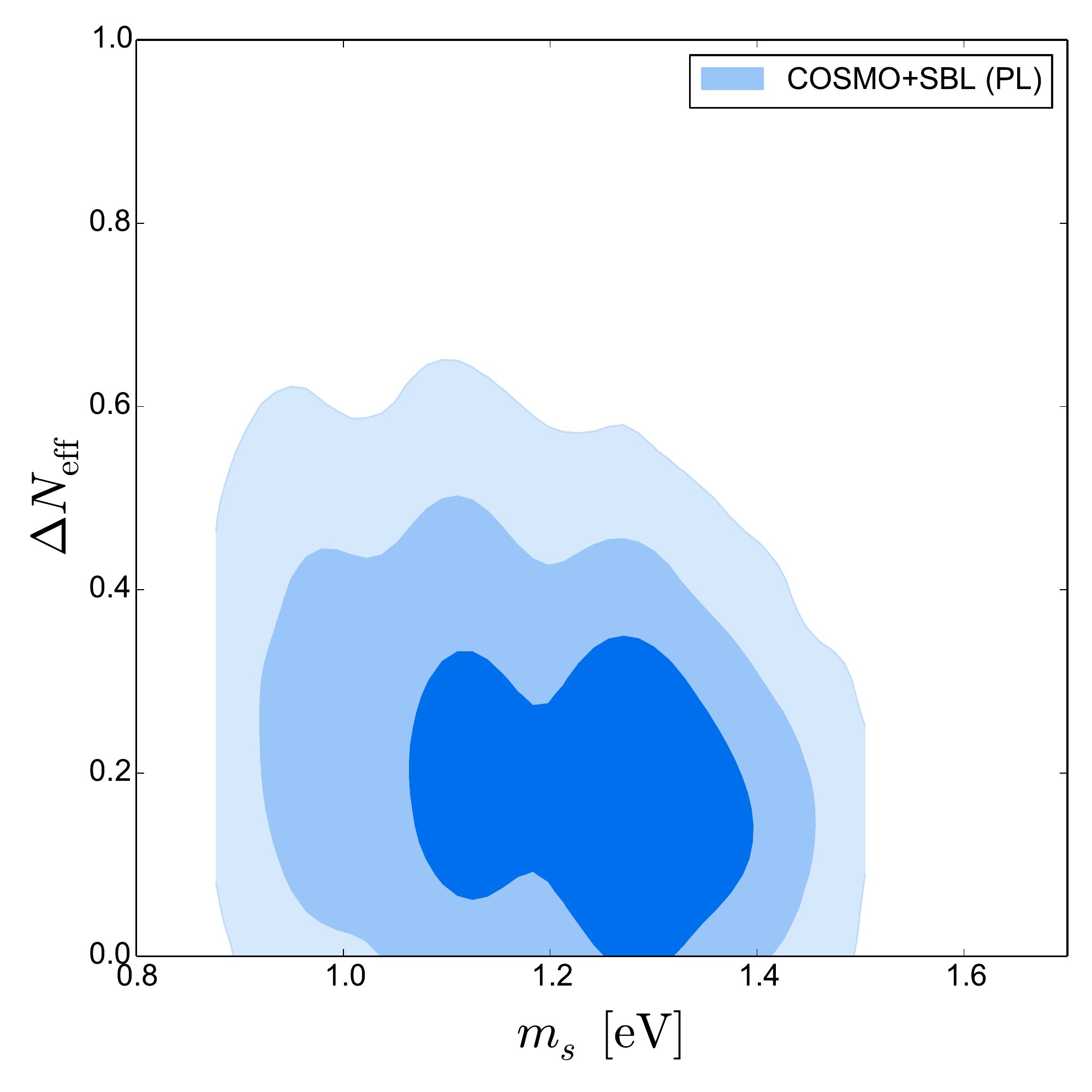}
\includegraphics[width=0.49\textwidth]{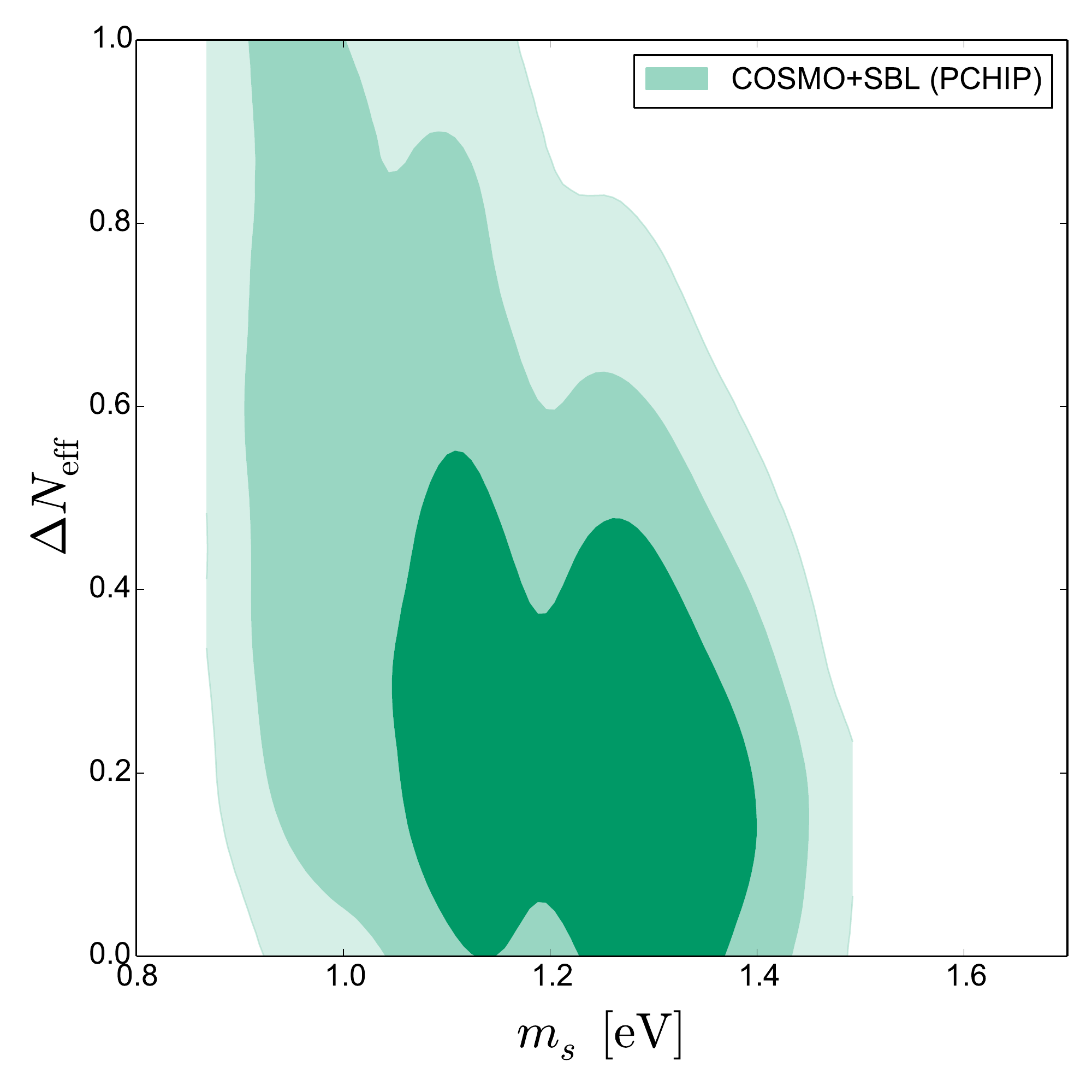}
\caption{\label{fig:posterior2DSBL}
$1\sigma$, $2\sigma$ and $3\sigma$ marginalized contours in the $m_s-\ns$ plane
in the fits with the SBL prior.
The left and right panels correspond, respectively,
to the standard power-law PPS and the \pchip PPS analyses.
}
\end{figure*}

Figure~\ref{fig:errorbars}
shows the marginalized $1\sigma$, $2\sigma$ and $3\sigma$ allowed intervals for
$\ns$ and $m_s$
that we obtained in the
COSMO(PL)
and
COSMO(PCHIP)
analyses,
without the SBL prior.
Figure~\ref{fig:posterior2DnoSBL} shows the
corresponding marginalized $1\sigma$, $2\sigma$ and $3\sigma$
allowed regions in the $m_s$--$\ns$ plane.
We can notice some major changes in the allowed values of both
$\ns$ and $m_s$
in the \pchip PPS case with respect to the power-law PPS case.
With a power-law PPS the best-fit value of $\ns$ is around 0.4,
whereas with the \pchip PPS it is at $\ns=1$,
that is the upper limit for $\ns$ assumed in the analysis.
The reason of this behavior is that
the effects of the presence of additional relativistic energy in the primordial universe
can be compensated by
an increase of the \pchip PPS at large $k$.
As a result,
the marginalized posterior for $\ns$
is increased in the region towards $\ns=1$,
in correspondence with higher values in the \pchip PPS for $k>0.35\mpcinv$.

Without the SBL constraint on $m_s$,
the different preferences for the value of $\ns$
in the power-law and \pchip PPS analyses
correspond to different allowed intervals for $m_s$.
As shown in Fig.~\ref{fig:errorbars},
although in both cases the
best-fit value of $m_{s}$ is near 0.6 eV,
the intermediate preferred region for $\ns$ in the power-law PPS analysis
gives for $m_{s}$ an upper limit of about 2 eV at $2\sigma$,
whereas the
large preferred values for $\ns$ in the \pchip PPS analysis
gives a tighter upper limit of about 1.5 eV at $2\sigma$.

The SBL prior on the sterile neutrino mass
$m_s$ puts a constraint so strong that in practice
the value of this parameter does not depend
on the inclusion or not of the freedom of the PPS.
In fact, the $m_s$ limits
in Tabs.~\ref{tab:plPPS} and \ref{tab:freePPS}
are similar in the power-law PPS and \pchip PPS analyses.
This can be seen also from the marginalized
allowed intervals of $m_s$ in
Fig.~\ref{fig:errorbars},
comparing the
COSMO+SBL(PL)
and
COSMO+SBL(PCHIP) allowed intervals.

A major difference occurs, instead, in the limits for $\ns$,
because
the effects of the presence of additional relativistic energy in the primordial universe
can be compensated by
an increase in the \pchip PPS at large $k$,
as in the case without the SBL constraint on $m_s$.
As shown in Fig.~\ref{fig:errorbars},
the best-fit and upper limits on $\ns$ in the
COSMO+SBL(PL)
and
COSMO+SBL(PCHIP)
are different.
In particular,
in the COSMO+SBL(PCHIP) the $3\sigma$ upper limit on $\ns$
allows the presence of a fully thermalized sterile neutrino compatible
with the SBL constraint on $m_s$.

Figure~\ref{fig:posterior2DSBL} shows the contour plots of
the marginalized $1\sigma$, $2\sigma$ and $3\sigma$ regions in the $m_s$--$\ns$ plane
that we obtained in the
COSMO+SBL(PL)
and
COSMO+SBL(PCHIP)
analyses.
The allowed regions in the left panel
are similar\footnote{
The only difference is that the analysis
in Ref.~\cite{Archidiacono:2014apa}
took into account also the
BICEP2 data
on the B-mode polarization of the CMB
\cite{Ade:2014xna}.
}
to those obtained in 
Ref.~\cite{Archidiacono:2014apa}
with a standard power-law PPS.
One can see that in this case
a fully thermalized sterile neutrino is quite disfavored.
On the other hand,
from the right panel
one can see that in the \pchip PPS analysis
a fully thermalized sterile neutrino with a mass just below $1\eV$ and with $\ns=1$ is even inside the $2\sigma$ region.
This means that a fully thermalized sterile neutrino can be accommodated in the cosmological model if the PPS is not forced to be described by a power-law.

\section{Best-fitting Primordial Power Spectrum}
\label{sec:results_PPS}

The results of our
\pchip PPS
analyses without and with the SBL prior on $m_{s}$
give interesting information on the shape of the PPS.

The marginalized posterior limits for the values $P_{s,j}$
in Eq.~(\ref{eq:pchip})
are listed in Tab.~\ref{tab:freePPS}.
One can see that
the least constrained nodes are the first and last, in $k_1$ and $k_{12}$,
for which there are only upper limits on the corresponding $P_{s,j}$.
This was expected,
because there are no data at the edges of
the wide interval of $k$
that we have considered.
For these two extreme nodes
the marginalized posterior is peaked on the lowest value
that we allowed in the fit (0.01).

On the contrary, the nodes from 5 to 10 are well constrained,
at the level of a few percent at $1\sigma$.
Considering the nodes from 2 to 4,
one can see that
the second node has preferred values higher than 1 within $1\sigma$,
the third node value is more than $2\sigma$ below 1 (around 0.6),
the fourth node value is again higher than 1 at more than $1\sigma$.
This implies that
the PPS that we obtained from the MCMC has a significant dip around $k_3\simeq0.002\mpcinv$ and
a less significant bump around $k_4\simeq0.0035\mpcinv$.

\begin{figure*}
\centering
\includegraphics[width=0.8\textwidth]{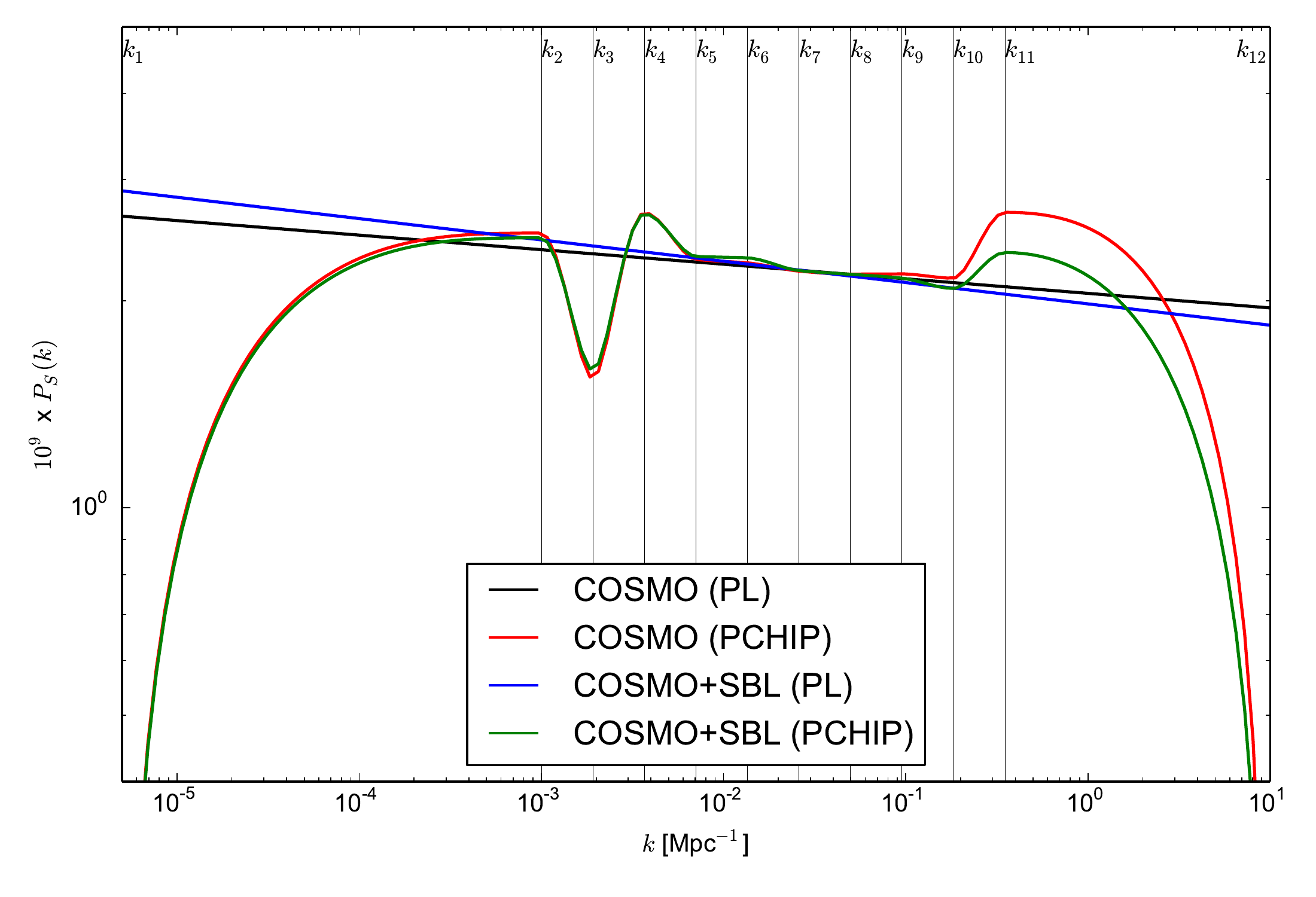}
\caption{\label{fig:bfPPS}
Best-fit PPS with different dataset combinations and PPS parameterizations.
}
\end{figure*}

To help the reader to visualize this feature,
we present in Fig.~\ref{fig:bfPPS} a comparison of the best-fitting PPS
\footnote{
We consider as the best-fitting PPS
that which
corresponds to the lower value
$\chi^2_{\text{min}}$
of
$\chi^2=-2\ln\mathcal{P}$,
where
$\mathcal{P}$
is the marginalized posterior probability in the space of the parameters
$P_{s,1}, \ldots, P_{s,12}$.
However,
one must take into account that in a parameter space with
a large number of dimensions $N_{\text{P}}$ the MCMC is not expected to explore
well the region near the true global best-fit corresponding to $\chi^2_{\text{min,true}}$.
In fact, the points are distributed mainly in a region where
$\chi^2 - \chi^2_{\text{min,true}} \sim N_{\text{P}}$.
Therefore,
the PPS that we consider as best-fitting
can be different from the true best-fitting PPS
in the intervals of $k$ which are not well constrained by the data.
}
in the power-law parametrization and in the \pchip parametrization,
without and with the SBL constraint.
One can see that
the best-fitting \pchip curves with and without the SBL prior
are significantly different only for $k \gtrsim k_{10}$.
The dip around $k_3\simeq0.002\mpcinv$ and
the bump around $k_4\simeq0.0035\mpcinv$
are clearly seen in the \pchip parametrization.

From Fig.~\ref{fig:bfPPS}
one can also see that
the \pchip parametrization has an approximate power-law behavior
between about $k_5\simeq0.007\mpcinv$ and $k_{10}\simeq0.2\mpcinv$.

\begin{figure*}
\centering
\includegraphics[width=0.9\textwidth]{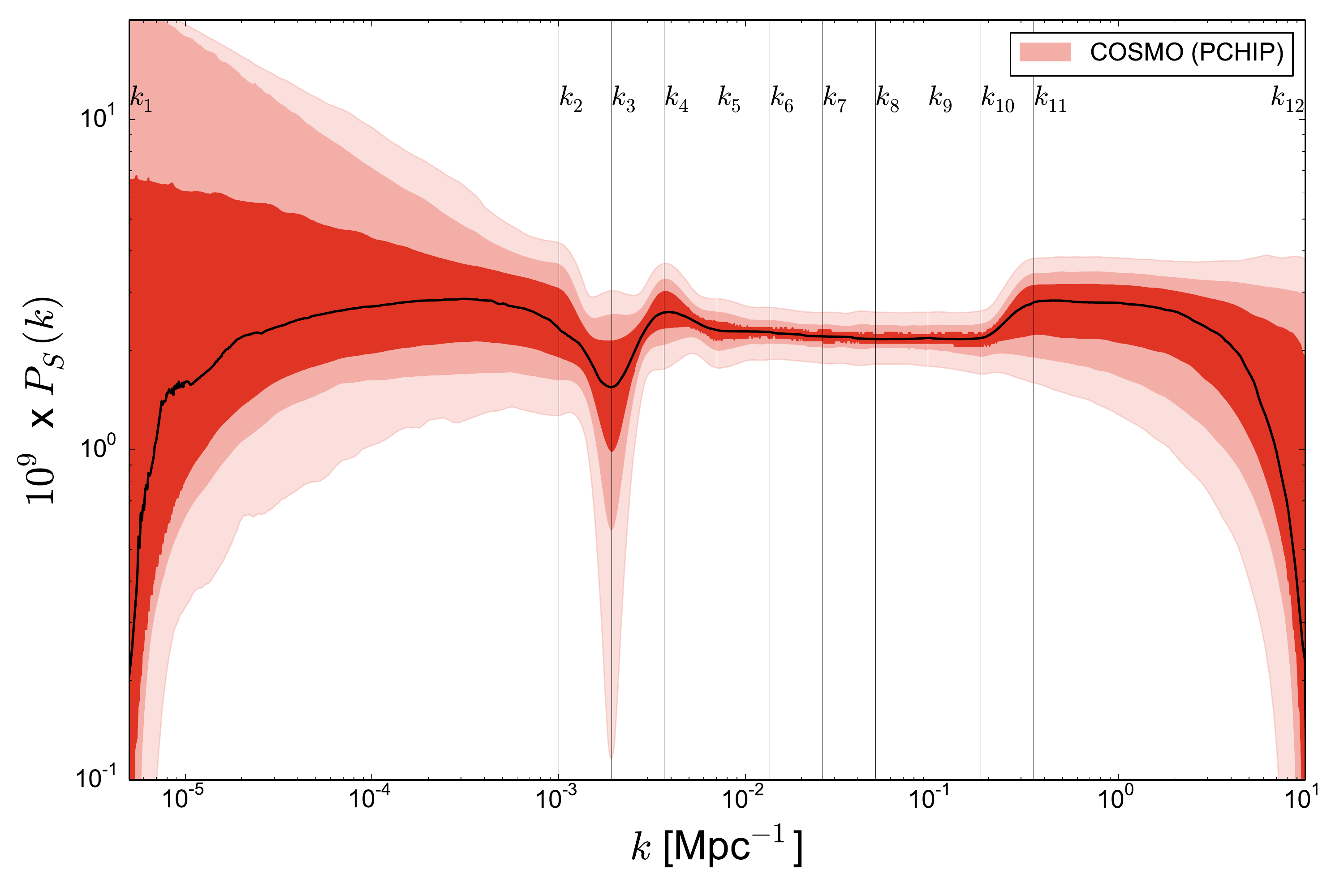}
\includegraphics[width=0.9\textwidth]{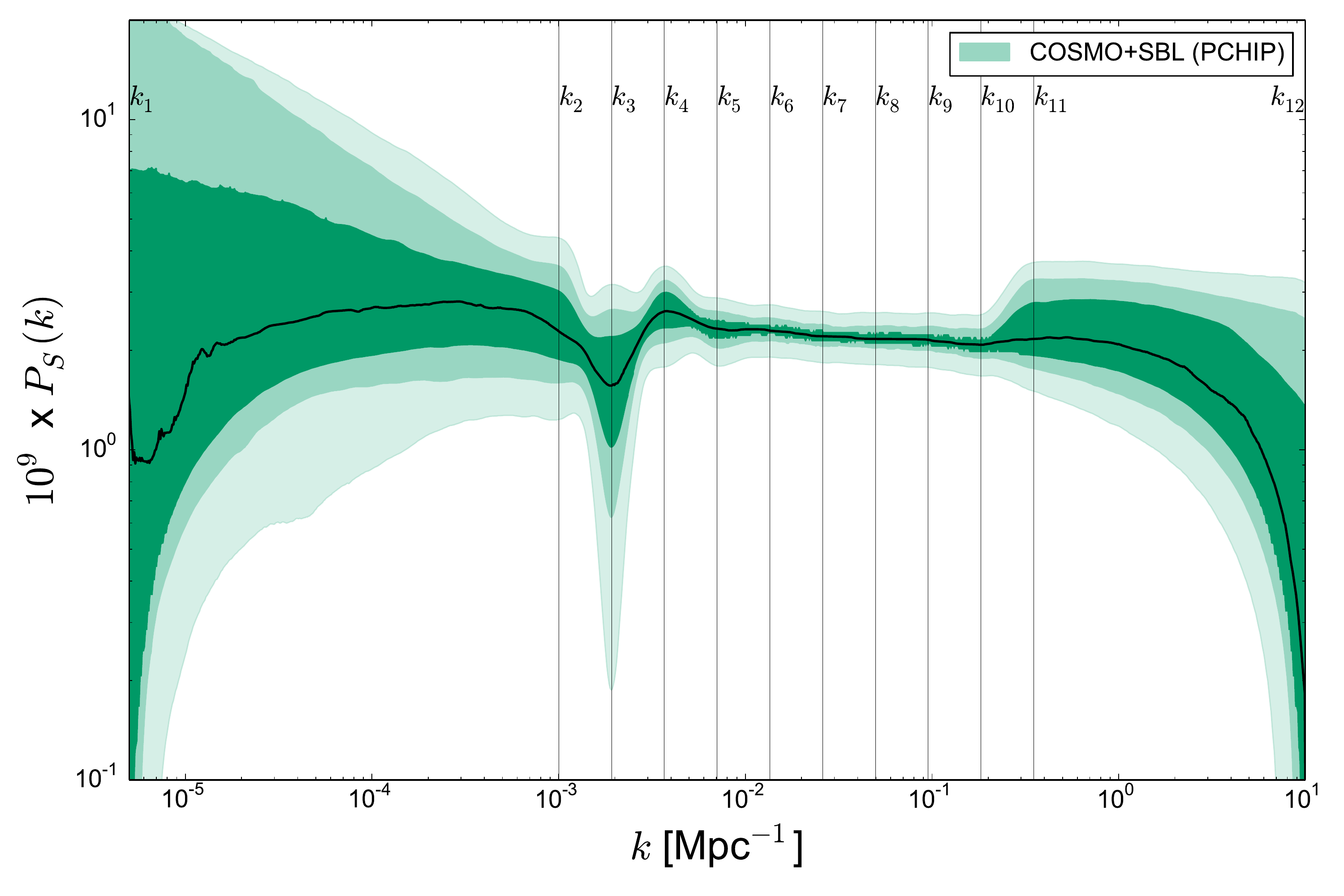}
\caption{\label{fig:bands}
Allowed
$1\sigma$,
$2\sigma$ and
$3\sigma$
bands of the
\pchip PPS obtained in the analyses
without (COSMO) and with (COSMO+SBL)
the SBL prior.
The bands have been obtained by marginalizing
the posterior distribution for each value of the wavenumber $k$
in a fine grid.
The black curves correspond to the maximum of the posterior distribution
for each value of $k$.
}
\end{figure*}

Another helpful way to visualize the behaviour of the
PPS obtained in the
analyses without and with the SBL prior
with the \pchip parametrization
is presented in Fig.~\ref{fig:bands},
which shows the
$1\sigma$,
$2\sigma$ and
$3\sigma$
bands obtained by marginalizing
the posterior distribution for each value of the wavenumber $k$
in a fine grid.
The two plots in Fig.~\ref{fig:bands}
show a well collimated band corresponding to the region in which the power-law gives a good approximation of the \pchip PPS,
between about $k_5\simeq0.007\mpcinv$ and $k_{10}\simeq0.2\mpcinv$.
Moreover,
the dip at $k\simeq0.002\mpcinv$ is well visible, as well as the bump at $k\simeq0.0035\mpcinv$.
On the other hand,
the widths of the bands diverge
for low and high values of $k$, where there are large uncertainties.

The major features that we have noticed in the reconstructed PPS
are in agreement with those found in Ref.~\cite{Hazra:2014jwa},
in which the scalar PPS has been reconstructed with a totally different technique,
the Richardson-Lucy iteration algorithm,
using the transfer function corresponding to the Planck 2013 best-fit for the $\Lambda$CDM model.
Apart for the suppression that they found around $k\simeq2\e{-4}\mpcinv$ and the features at higher $k$, the main differences with respect to the power-law PPS are the same that we found in our analysis.
According to the authors of Ref.~\cite{Hazra:2014jwa},
these major features are related to the low-$\ell$ spectrum of the temperature perturbations measured by the Planck experiment, that obtained a dip in the power around $\ell\simeq22$ and a slight excess around $\ell\simeq40$.

Although the parametrization with a natural cubic spline is noisy due to spurious oscillations
between the nodes,
also in Fig.~8 of Ref.~\cite{dePutter:2014hza} it is possible to guess the presence of a dip around $k\simeq0.002\mpcinv$,
especially in the curves with more than 20 nodes.
However,
our parametrization is much cleaner and permits a better visualization of these features.

\section{Conclusions}
\label{sec:disc}

In this work we analyzed the effects of a free form of the primordial power-spectrum
of scalar fluctuations,
which is not constrained to the usual power-law form
that is predicted by the simplest models of inflation
(see Refs.~\cite{Lyth:1998xn,Bassett:2005xm,Baumann:2008bn}).
This freedom in the PPS could arise from a more complicate inflationary mechanism
(see Refs.~\cite{Martin:2014vha,Kitazawa:2014dya} and references therein).

We parameterized the PPS with a
``piecewise cubic Hermite interpolating polynomial''
(\pchip)
described in details in
Appendix~\ref{sec:ppsparametrization}.
Our \pchip parameterization of the PPS depends from
the values of the PPS in twelve nodes
(given in Eq.~(\ref{eq:nodesspacing}))
which cover a wide range of values of the wavenumber $k$.
We choose the \pchip method
in order to avoid spurious oscillations of the interpolated function between the nodes
that can be obtained with spline interpolations
(see Refs.~\cite{Hlozek:2011pc,dePutter:2014hza}).

We performed an analysis of cosmological data in which only the primordial spectrum of scalar perturbations is considered,
neglecting the controversial
\cite{Ade:2013zuv,Adam:2014bub}
data on the B-mode polarization of the CMB
\cite{Ade:2014xna}
which would require to take into account also the primordial spectrum of tensor perturbations.
We used the most precise CMB measurements together with low-redshift measurements of the Hubble parameter, the galaxy distribution and the matter distribution in the Universe
(see Section~\ref{sub:cosmodata}).

We found that
the freedom of the form of the PPS does not affect significantly the
fitted values of the parameters in the $\Lambda$CDM model,
while the results concerning the existence of a sterile neutrino in the early Universe
can change drastically.
If we do not impose any prior on
the sterile neutrino mass $m_s$ from the results of short-baseline oscillation experiments
(see Section~\ref{sub:numix}),
a larger value for the sterile neutrino contribution
$\ns$
to the effective number of relativistic degrees of freedom
before photon decoupling
is preferred in the \pchip PPS parameterization
with respect to the standard power-law parameterization.
The marginalized best fit of $\ns$ is moved towards one,
which corresponds to a fully thermalized sterile neutrino.
This shift corresponds to a tightening of the cosmological preferred values for $m_s$.

In the analysis with a prior on $m_s$ obtained from the fit of
short-baseline oscillation experiments
\cite{Giunti:2013aea},
the freedom of the \pchip PPS affects only the bound on $\ns$,
because the allowed range of $m_s$
is strongly constrained by the SBL prior.
We found that a free form of the PPS allows the existence in the early Universe of
a fully thermalized sterile neutrino with a mass of about 1 eV
\cite{Dolgov:2003sg,Cirelli:2004cz}.
This possibility is quite disfavored by the analysis of cosmological data
with a power-law PPS
\cite{Valentino:2013wha,Archidiacono:2013xxa,Mirizzi:2013kva,Gariazzo:2013gua,Archidiacono:2014apa,Bergstrom:2014fqa}.
Hence,
the freedom of the PPS allows us to reconcile the cosmological data with
short-baseline neutrino oscillations
without the need of an additional mechanism which suppresses the
thermalization of the sterile neutrino
\cite{Chu:2006ua,Hannestad:2012ky,Mirizzi:2012we,Saviano:2013ktj,Hannestad:2013pha,Hannestad:2013ana,Dasgupta:2013zpn,Bringmann:2013vra,Ko:2014bka,Archidiacono:2014nda,Saviano:2014esa,Mirizzi:2014ama,Rehagen:2014vna,Ho:2012br}.

We obtained also some interesting information on the form of the PPS,
whose behavior is well constrained by the cosmological analysis for
$0.001\mpcinv \lesssim k \lesssim 0.3\mpcinv$.
In particular,
we have shown that in the range
$0.007\mpcinv \lesssim k \lesssim 0.2\mpcinv$
the PPS can be approximated with a power-law
and the values of the PPS in the nodes of the \pchip parameterization
lying in this range of $k$
have only a few-percent uncertainty.
The PPS in the range
$0.001\mpcinv \lesssim k \lesssim 0.0035$ presents a clear
dip at $k \simeq 0.002\mpcinv$, with a statistical significance of more than 2$\sigma$,
and a small bump at $k \simeq 0.0035\mpcinv$, with a statistical significance of about $1\sigma$.
These features of the PPS are in agreement with
those found in Ref.~\cite{Hazra:2014jwa} with a completely different method.

In the future the analysis presented in this work could be repeated with the inclusion of a parametrization for the primordial spectrum of tensor perturbations
when improved data on the B-mode polarization of the CMB
will be available.
This will allow us to study with more precision the few relics of cosmological inflation that we can access.

\begin{acknowledgments}
We would like to thank
M. Archidiacono,
E. di Valentino,
N. Fornengo,
S. Hannestad,
A. Melchiorri,
Y.F Li
and
H.W. Long
for stimulating discussions
and fruitful collaboration in previous works.
This work is supported by the research grant {\sl Theoretical Astroparticle Physics} number 2012CPPYP7 under the program PRIN 2012 funded by the Ministero dell'Istruzione, Universit\`a e della Ricerca (MIUR).
\end{acknowledgments}

\appendix
\section{\texttt{PCHIP} Parametrization of the Primordial Power Spectrum}
\label{sec:ppsparametrization}

In this work we parameterized the PPS with a
``piecewise cubic Hermite interpolating polynomial''
(\pchip)
\cite{Fritsch:1980,Fritsch:1984}.
We decided to adopt this interpolating function in order to avoid
spurious oscillations of the interpolating function between the nodes
which is often obtained in spline interpolations.
This problem occurs because a natural cubic spline requires the values of the function, the first and the second derivatives to be continuous in the nodes \cite{NR}.

The \pchip function, instead, is constructed in order to preserve the
shape of the set of points to be interpolated.
This is achieved with
a modification of the ``monotone piecewise cubic interpolation''
\cite{Fritsch:1980}
which can accommodate non-monotone functions
and preserves the local monotonicity.

Let us consider a function with known values $y_{j}$
in N nodes $x_{j}$,
with $j=1,\ldots,N$.
A piecewise cubic interpolation is performed with
$N-1$ cubic functions between the nodes.
The determination of these
$N-1$ cubic functions requires the determination of
$4(N-1)$ coefficients.
Besides the
$2(N-1)$ constraints obtained by requiring that the initial and final
point of each cubic function match the known values of the original function
in the corresponding nodes,
one needs a prescription for the other $2(N-1)$ necessary constraints.
In the case of a natural cubic spline interpolation one gets
$2(N-2)$ constraints by requiring the continuity of the first and second derivatives in the nodes
and the remaining two constraints
are obtained by requiring that the second derivatives in the first and last nodes vanish.
The drawback of this method is that the interpolating curve
is determined by a set of linear equations without any local control.
In fact,
all the interpolating curve is affected by the change of a single point.

\begin{figure*}
\centering
\includegraphics[width=0.8\textwidth]{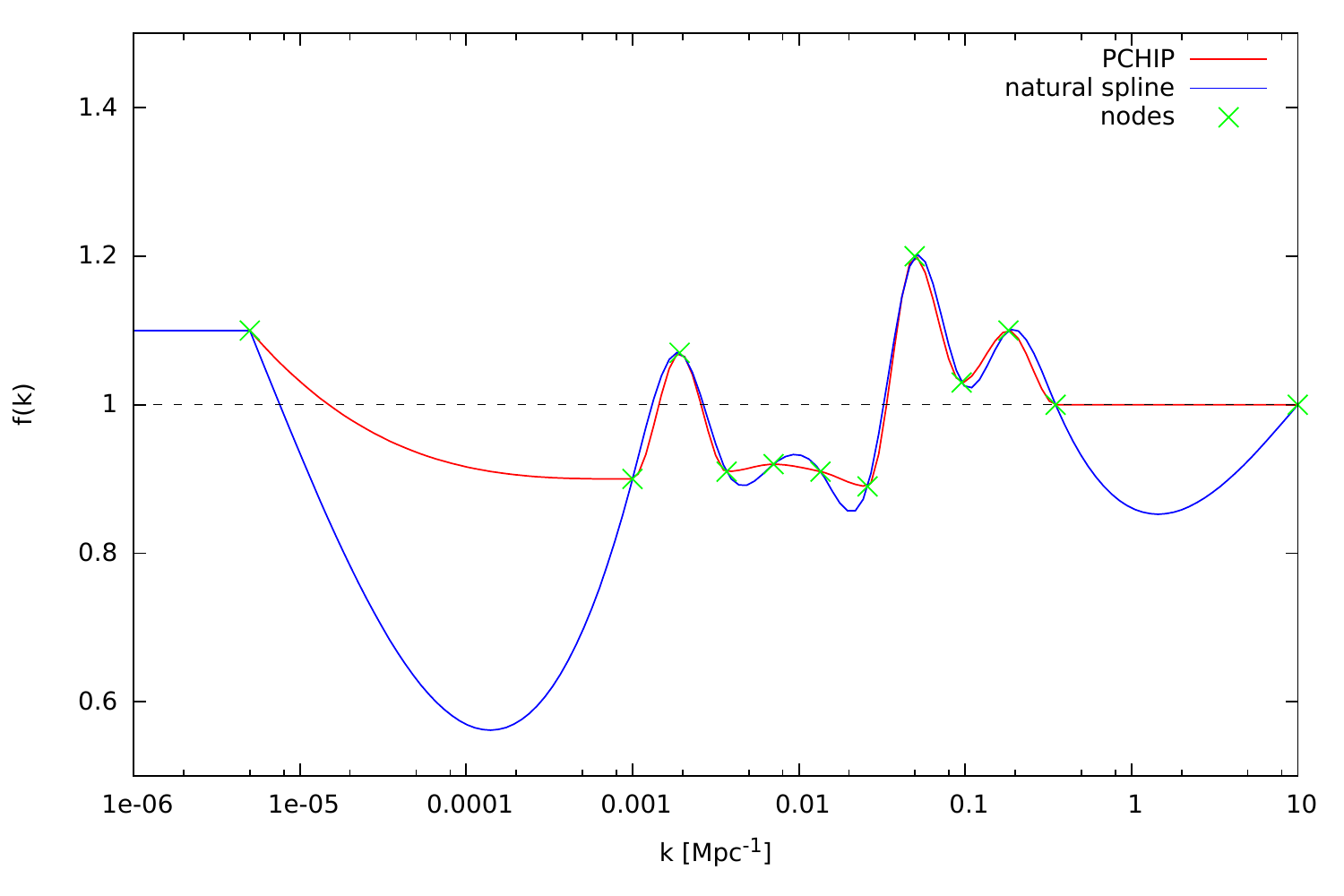}
\caption{\label{fig:interpolation}
Illustration of the difference between the \pchip (red line) and the natural spline (blue line) interpolations
$f(\log{k}; y_{1}, \ldots, y_{12})$
of a function with known values
$y_{1}, \ldots, y_{12}$
in 12 nodes (green crosses) at the values of $k$ in Eq.~\ref{eq:nodesspacing}.
The values $y_{1}, \ldots, y_{12}$ in the nodes are
1.1,
0.9,
1.07,
0.91,
0.92,
0.91,
0.89,
1.2,
1.03,
1.1,
1.0,
1.0.
}
\end{figure*}

Local control of the interpolating curve can be achieved by
relaxing the requirement of continuity of the second derivatives
in the nodes
and using the resulting freedom to adjust the first derivatives
with a local prescription.
In order to see how it can be done, it is convenient to write the cubic interpolating polynomial
between the nodes $x_{j}$ and $x_{j+1}$ in the Hermite form
\begin{equation}
f(x; y_{1}, \ldots, y_{N})
=
\frac{\left( h_{j} + 2 t \right) \left( h_{j} - t \right)^2}{h_{j}^3} y_{j}
+
\frac{\left( 3 h_{j} - 2 t \right) t^2}{h_{j}^3} y_{j+1}
+
\frac{\left( h_{j} - t \right)^2 t}{h_{j}^2} d_{j}
+
\frac{t^2 \left( h_{j} - t \right)}{h_{j}^2} d_{j+1}
,
\label{a01}
\end{equation}
where
$t = x-x_{j}$
and
$h_{j} = x_{j+1}-x_{j}$.
Here
$d_{j}$ and $d_{j+1}$
are the values of the derivatives in the two nodes.
In the \pchip method the derivatives are chosen in order to preserve the local
monotonicity of the interpolated points.
This is done by considering the relative differences
\begin{equation}
\delta_{j}
=
\frac{y_{j+1} - y_{j}}{x_{j+1}-x_{j}}
.
\label{a02}
\end{equation}
The \pchip  prescription is:
\begin{itemize}

\item
If $\delta_{j-1}$ and $\delta_{j}$ have opposite signs,
then $x_{j}$ is a discrete local minimum or maximum
and
$d_{j}=0$.

\item
If $\delta_{j-1}$ and $\delta_{j}$ have the same sign,
then
$d_{j}$
is determined by the weighted harmonic mean
\begin{equation}
\frac{w_{1}+w_{2}}{d_{j}}
=
\frac{w_{1}}{\delta_{j-1}}
+
\frac{w_{2}}{\delta_{j}}
,
\label{a03}
\end{equation}
with
$w_{1}=2h_{j}+h_{j-1}$
and
$w_{2}=h_{j}+2h_{j-1}$.

\item
The derivatives in the first and last nodes
are determined by a shape-preserving prescription
based on a quadratic fit of three points.
For $d_{1}$ we consider the three points
$(x_{1},y_{1})$,
$(x_{2},y_{2})$,
$(x_{3},y_{3})$.
The derivative in $x_{1}$ of the parabola which passes through these three points is
given by
\begin{equation}
d(h_{1}, h_{2}, \delta_{1}, \delta_{2})
=
\frac{\left( 2 h_{1} + h_{2} \right) \delta_{1} - h_{1} \delta_{2}}{h_{1} + h_{2}}
\,.
\label{a04}
\end{equation}
The shape-preserving prescription for $d_{1}$ is:

\begin{itemize}

\item
If the signs of $d(h_{1}, h_{2}, \delta_{1}, \delta_{2})$
and $\delta_{1}$ are different,
then
$d_{1} = 0$.

\item
If the signs of $\delta_{1}$ and $\delta_{2}$ are different
and
$|d(h_{1}, h_{2}, \delta_{1}, \delta_{2})| > 3 |\delta_{1}|$,
then
$d_{1} = 3 \delta_{1}$.

\item
Else $d_{1} = d(h_{1}, h_{2}, \delta_{1}, \delta_{2})$.

\end{itemize}

For $d_{N}$ one must replace $1 \to N-1$ and $2 \to N-2$.

\end{itemize}

We fit the power spectrum
$P_{s}(k)$
with Eq.~(\ref{eq:pchip}),
in which the function
$\pchip(k; P_{s,1}, \ldots, P_{s,12})$
is calculated with the \pchip prescription
in the logarithmic scale of $k$:
\begin{equation}
\pchip(k; P_{s,1}, \ldots, P_{s,12})
=
f(\log{k}; P_{s,1}, \ldots, P_{s,12})
\,.
\label{a11}
\end{equation}

A comparison between the natural cubic spline and the \pchip interpolations
of the PPS is presented in Fig.~\ref{fig:interpolation}.
We choose the same nodes positions that we used for the PPS parametrization in our cosmological analysis
and we choose the values of the function in the nodes
in order to show the difference between
the natural cubic spline and the \pchip interpolations.
One can see that
the \pchip interpolation can reproduce the shape of the points without adding
the spurious features between the points that are clearly visible
in the natural cubic spline interpolation.

%

\bibliographystyle{utcaps}

\bibliography{pps}


\end{document}